\documentclass[flushrt,preprint]{aastex}       
\input{psfig.tex}  
\begin{document}

\newcommand{\ds}{$\Delta {\rm S}\,$}
\newcommand{\msun}{M$_\odot$}

\slugcomment{Astronomical Journal, in press}

\shorttitle{Metallicity of RRd's in the LMC}
\shortauthors{Bragaglia et al.}

\title{METALLICITIES FOR DOUBLE MODE RR LYRAE IN THE LARGE MAGELLANIC CLOUD
\footnote{Based on observations collected at the European Southern 
 Observatories, Chile} }

\author{
A. Bragaglia\altaffilmark{1}, 
R.G. Gratton\altaffilmark{2}, 
E. Carretta\altaffilmark{2}, 
G. Clementini\altaffilmark{1}, 
L. Di Fabrizio\altaffilmark{1}, 
M. Marconi\altaffilmark{3}}
\altaffiltext{1}{Osservatorio Astronomico di Bologna, Via Ranzani 1, 
  40127 Bologna, Italy}
\altaffiltext{2}{Osservatorio Astronomico di Padova, Vicolo 
 dell'Osservatorio 5, 35122 Padova, Italy}
\altaffiltext{3}{Osservatorio Astronomico di Capodimonte, Via Moiariello 16, 
I-80131 Napoli, Italy}
\email{angela@bo.astro.it, gratton@pd.astro.it, carretta@pd.astro.it, 
 gisella@bo.astro.it, lau@bo.astro.it, marcella@na.astro.it}

\begin{abstract}

Metallicities for six double mode RR Lyrae's (RRd's) in the Large Magellanic
Cloud have been estimated using the \ds ~method. The derived [Fe/H] values
are in the range [Fe/H] = $-$1.09 to $-$1.78 (or $-0.95$ to $-$1.58, adopting 
a different calibration of [Fe/H] $vs$ \ds).  
Two stars in our sample are at the very metal rich limit of all RRd's  for
which metal abundance has been estimated, either by direct measure (for field
objects) or on the basis of the hosting system (for objects in globular
clusters or external galaxies).

These metal abundances, coupled with mass determinations from pulsational
models and the Petersen diagram, are used to compare the mass-metallicity
distribution of field and cluster RR Lyrae variables.

We find that field and cluster RRd's seem to follow the same
mass-metallicity distribution, within the observational errors, strengthening
the case for uniformity of properties between field and cluster variables

At odds to what is usually assumed, we find no significative difference in
mass for RR Lyrae's in globular clusters of different metallicity and
Oosterhoff types, or there may even be a difference contrary 
to the commonly accepted one,
depending on the metallicity scale adopted to derive masses. This ``unusual'' 
result for the mass-metallicity relation is probably due, at least in 
part,  to  the inclusion of updated opacity tables in 
the computation of metal-dependent pulsation models.

\end{abstract}


\newpage

\section{Introduction}

The well-known dichotomy between {\it short} and  {\it long} distance scales
derived from old, population II stars has plagued astronomers for a long time
and not even the impressive improvements in measuring distances due to the
Hipparcos mission (see e.g., Gratton et al. 1997) seemed to provide an
universally accepted solution.
This is pointed out by the disagreement
existing between results based on field or cluster stars.
Methods founded on field stars, like statistical parallaxes or the
Baade-Wesselink method applied to field RR Lyrae's, or direct analysis of
field horizontal branch stars (see Gratton 1998) seem to favour the {\it
short} scale. Instead, cluster star-based distances, derived from
main-sequence fitting to local subdwarfs (e.g. Gratton et al. 1997), or from
pulsational properties of RR Lyrae's in globular clusters 
(Sandage 1993),
or from the calibration of the HB luminosity level using the Cepheid distance
modulus of the LMC (Walker 1992), seem to support the {\it long} scale.

Following a calibration of the absolute magnitude of horizontal branch field
metal-poor stars with good parallaxes from Hipparcos, Gratton (1998) suggested
that a difference (at a $\sim$ 0.1--0.2 mag level) might actually exist
between the luminosity of HB stars in globular clusters (GC's) 
and in the field. This
result appeared to be confirmed by the difference in the average magnitude for
field and cluster RR Lyrae's in the LMC measured by Alcock et al. (1996) and
Walker (1992) respectively (but see Clementini et al. 2001,
hereafter C2001).
Another indication in favour of this possible difference comes from the latest
evolutionary models by Sweigart (1997), which include some extra-mixing of He
and other heavy elements (C, N, O): as a matter of fact there is observational
evidence for extra-mixing in cluster red giants but not among field stars
(Gratton et al. 2000). On the other side, studies by Catelan (1998) and by
Carretta, Gratton \& Clementini (2000a) put this suggestion to a test using
the pulsational properties of a selected sample of field and cluster RR
Lyrae's. Carretta et al. (2000a) show in a quantitative way that in our Galaxy,
field and cluster RR Lyrae's cannot be distinguished in a $\Delta \log P_{\rm
(field-cluster)}$ -- [Fe/H] plane: this goes towards excluding the possibility
of a luminosity difference.
In fact, the pulsational Period-Mass-Luminosity-$T_{\rm eff}$ relation (van
Albada \& Baker 1971) tells us that, at fixed $T_{\rm eff}$, a difference in
the period distribution (or its absence) indicates that there is (or there is
not) a difference in the M/L ratio for the variables. The determination of
star masses with the high precision required to settle the question is still
one of the  most difficult tasks in astrophysics. In this context double mode
RR Lyrae's are extremely powerful tools, because their masses can be estimated
from the ratio of the two pulsational periods using pulsational models, hence
independently from stellar evolution models.
Unluckily, the $M-$[Fe/H] relation can presently be studied in our Galaxy only
using double mode pulsators in {\it clusters}, since only a handful of RRd's
have been so far identified in the {\it field} of our Galaxy.

The field RR Lyrae's in the Large Magellanic Cloud bar play a key role in this
respect, since i) projection effects are negligible, and
they can be considered  at the same distance from us 
and, ii) plenty of double pulsating variables have been
identified in the field of the LMC by the MACHO experiment (Alcock et al.
1997, hereinafter A97). Field RR Lyrae's in the LMC with good [Fe/H]
determinations may allow the derivation
of both an accurate ${\rm L}=f$([Fe/H]) (using
single mode pulsators) and a ${\rm M}=f$([Fe/H]) relation, if any, (using
double mode pulsators) within a homogeneous sample of field variables. The
accuracy required to settle the question of the magnitude difference can be
estimated from the pulsation equation of van Albada \& Baker (1971), computed
at fixed temperature and period: in order to appreciate a supposed difference
of $\Delta \log{\rm L}\sim$ 0.05 (or 0.12 mag) we must detect a $\Delta
\log{\rm M}\sim 0.06$\ (i.e. $\sim 0.12~{\rm M}_\odot$) difference in mass
between field and cluster variables.
Cox (1991) using a fixed metallicity and the new OPAL (Rogers \& Iglesias 1992)
opacity tables found masses of 0.65 \msun ~for RR Lyrae pulsators in
Oosterhoff type I clusters (Oo I; Oosterhoff 1939), and of 0.75 to even 0.85 
\msun  ~in Oosterhoff type II clusters (Oo II). Since 
the approximate difference in metallicity between the two Oo type clusters 
he studied is 
0.5-0.7 dex, we estimate that an accuracy $\leq 0.2$~dex in [Fe/H] is then 
required.

We started an observational program on the field RR Lyrae's of the LMC i) to
derive their average apparent luminosity, and ii) to determine their
metallicities (in particular for the RRd pulsators). The first part of the
program, mostly dealing with new photometric data acquired by our team  near
the bar of the LMC, is described in C2001 and Di Fabrizio et al. (2001).
The main
results are the very accurate determination of the mean apparent magnitude of
the field RR Lyrae's of the LMC and the
derivation of an independent reddening estimate.

From the light curves so derived we determined epochs to properly time the 
spectroscopic observations.
In this paper we present metallicities with an accuracy of $\Delta$ [Fe/H]
$\sim 0.2$ dex for six RRd's in the LMC. We used the \ds ~index (Preston
1959), defined as \ds = 10[SpT(H)--SpT(K)], where SpT(H) and SpT(K) are the
spectral types measured at minimum light, in units of tenths of a spectral
type class, based on the hydrogen lines and on the CaII K line, respectively.
In Section 2 we present the observational data and in Section 3 we describe
the derivation of the \ds ~values for the 6 stars. Section 4 is devoted to the
determination of metallicities; in Section 5 we derive masses of our targets
from the Petersen diagram, and discuss the mass-metallicity distribution.
Summary and conclusions are presented in Section 6.

\section{Observations and Reductions}

\begin{figure}
\vspace{12cm}
\includegraphics{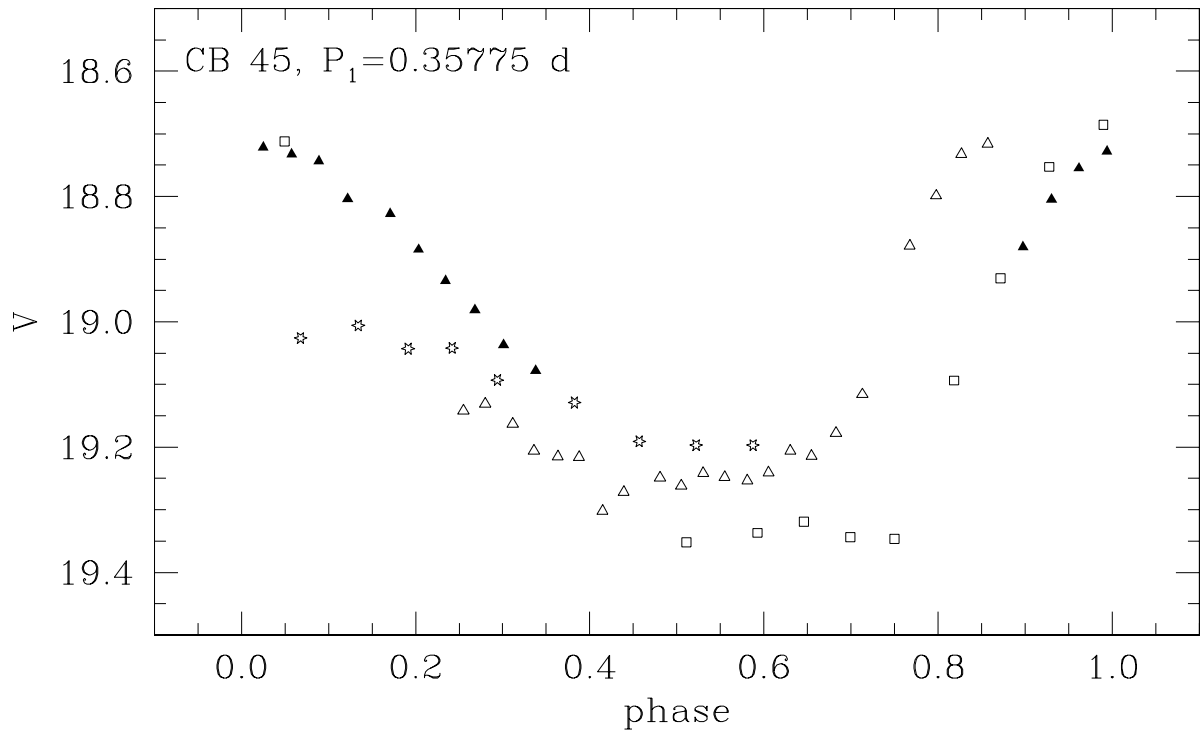}
\caption{The LMC RRd CB 45: V light curve derived from C2000
photometry; the different symbols refer to the 4 different nights of
observation. Data have been phased
according to the first overtone period of pulsation (P$_1$) given in A97 and
the epoch derived by C2000.\label{fig-cb45}}
\end{figure}

\begin{figure}
\vspace{15cm}
\includegraphics{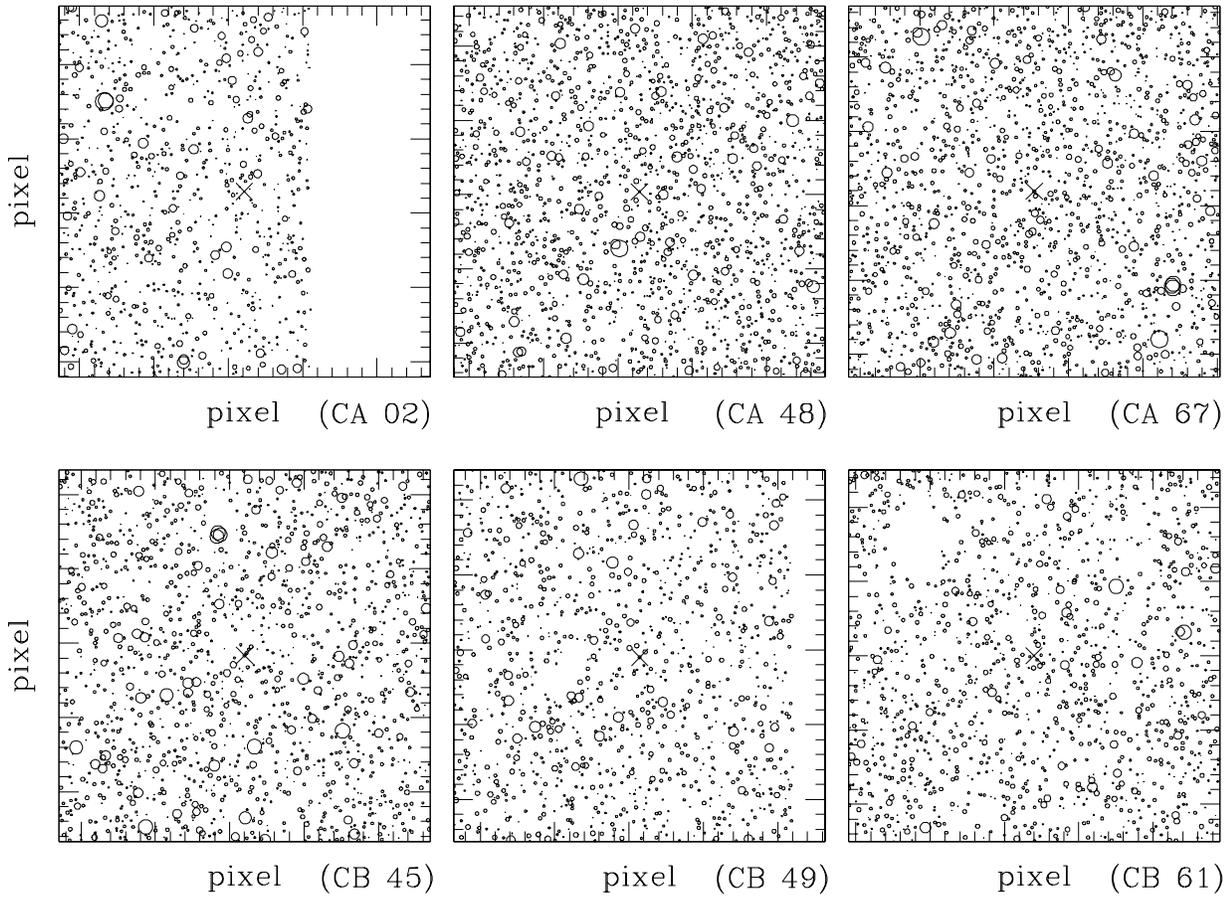}
\caption{Finding charts for the 6 program stars derived from
C2000 photometry (at top, from left to right: CA 02, CA 48, CA 67; at bottom,
from left to right: CB 45, CB 49, CB 61). Each RRd is at the center,
indicated by the cross (coordinates are found in Table 1);
the fields shown cover 200 arcsec$^2$, with North at top and East on the left.
\label{fig-map}}
\end{figure}

The MACHO program discovered about 7,900 RR Lyrae's in the twenty-two 40
arcmin$^2$ fields centered on the LMC bar (Alcock et al. 1996), and published
coordinates and periods (but no epochs) for 73 RRd's in the LMC bar (A97; 
differential light curves can be found in the MACHO website, e.g. 
\url{wwwmacho.mcmaster.ca}). 
Since the \ds ~method requires the acquisition of spectra when variables are
at minimum light, new photometric observations were obtained in order to
derive complete ephemerides for some of A97 double mode pulsators. All
observations were carried out in La Silla, Chile, in January 1999.
Four nights at the Danish 1.54m telescope, one of which of photometric
quality, were dedicated to Johnson V and B photometry (see C2001 for a
complete description) and immediately reduced to produce light curves and
derive ephemerides (epochs, in particular) for our targets. We tried to
maximize the number of RRd's observable in a single DFOSC (Danish Faint
Object Spectrograph and Camera) pointing (field of
view 13.5 arcmin$^2$), and for the present program we selected 2 positions
(in MACHO fields \#6 and \#13) comprising a total of 9 RRd's; we refer to
them as ``field A''
($\alpha_{2000}=05^h 22^m 44^s$ $\delta_{2000}=-70^o 34' 15"$) 
and ``field B''
($\alpha_{2000}=05^h 17^m 28^s$ $\delta_{2000}=-71^o 00' 14"$); 
henceforth RRd's in the two fields will be indicated by CA or CB followed by an
ordering number derived from tab. 1 of A97. Periods in A97 were used together
with the epochs obtained from the new photometric data to properly schedule at
minimum light the subsequent spectroscopic observations, done on 2 nights only
10 days afterwards.
We did not use our own derived periods, since C2001 sampling of the light
curve of the double mode pulsators (about 60 data-points in 4 consecutive
nights) does not allow the derivation of periods for the RRd's as accurate as
in A97.
We show in Figure~\ref{fig-cb45} the light curve for one of the RRd's based on
C2001 data, and phased using A97 first overtone period P$_1$. Finding charts
for the 6 stars here discussed are presented in Figure~\ref{fig-map}, taken
from C2001 photometry; maps are 200 arcsec$^2$, and the position of each RRd
is indicated by a cross (not in scale), right at the center of the fields.

The spectroscopic observations were carried out at the 3.6m ESO telescope,
during the nights 1999 January 17-18. EFOSC2 (ESO Faint Object Spectrograph
and Camera) was used, mounting the CCD \#40,
a Loral $2k\times 2k$\ chip, binned $2\times 2$, in combination with grism \#7
(600 lines/mm, 3270-5240 \AA) and a 1.5 arcsec wide slit, resulting in a
$\simeq$ 9~\AA ~resolution (or R$\simeq$450). Whenever possible, the slit was
rotated in order to exclude nearby contaminating stars. Sky conditions were
good enough for spectroscopic observations, but not of photometric quality,
and the seeing varied from about 1" to 1.9", with an average value around
1.5".

We observed close to minimum light 7 of the 9 RRd's within these fields,
exploiting the previously described ephemerides, obtaining usable spectra for
6 of them (for one of the stars the spectrum is heavily contaminated by
a very close bright object falling into the slit, and we excluded it).
Identifications and information on these 6 targets are provided in
Table~\ref{tab-rrd1}. Equatorial coordinates come from the transformation
of pixel positions to right ascension and declination, based on about one
hundred stars individuated both in each of the two fields and on the
Digitized Sky Survey\footnote{
The Digitized Sky Survey was produced at the Space Telescope
Science Institute under U. S. Government grant NAG W-2166.};
they differ only a few arcsec from the ones published in A97.
The $<{\rm V}>$ values are intensity averages 
and should be taken as preliminary, since a new calibration of
the photometric data is under way (see C2001);
epochs of maximum light are given in column 8.
The fundamental and first overtone periods P$_0$ and P$_1$ are taken from
A97. In column 9 and 10 we give the Heliocentric Julian Day of the
spectroscopic observations of our targets and the corresponding phases
computed using A97 first overtone pulsation periods and the epochs in column 8.
Given the faintness of our targets (V$\simeq$19.5 at minimum light), exposure
times were in the range from 30 to 50 minutes, in order to reach the maximum
possible S/N and avoid phase blurring on the pulsation cycle. Typical S/N
ratios range from 10 to 30.

Thirteen non variable stars in the open cluster Collinder 140 (Cr 140,
Clari\'a \& Rosenzweig 1978, hereinafter CR), observed with the same
instrumental configuration, were chosen as spectral type standard stars, to
apply the "classical" approach to the \ds ~method via spectral types. The
observed stars are indicated in Table~\ref{tab-cr140}, together with their
Jonhson B, V and  Str\"omgren $b-y$ photometry, and spectral types derived
both from CR, and from the Str\"omgren $b-y$ photometry (see next Section).
The last column of the table gives the spectral type adopted in the present
analysis: CR values or the average of CR types with those derived from the
$b-y$, when both estimates are available.

For calibration purposes (i.e. to check that our $\Delta$S's are on the
standard system) we took spectra at minimum light of 6 field $ab$ type RR
Lyrae's of known \ds, namely IU Car, X Crt, WY Ant, AF Vel, U Lep and TV Leo;
phases were derived from published ephemerides, and checked
a posteriori by the spectral types (see note to Sect. 3.1).
We also acquired 10 spectra along the pulsation cycle (from phase 0.0 to
0.65) of one $c$~type RR Lyr (T Sex, also of known \ds), in order to derive
phase corrections for the spectra not exactly taken at minimum light.

\begin{figure}
\vspace{18cm}
\includegraphics{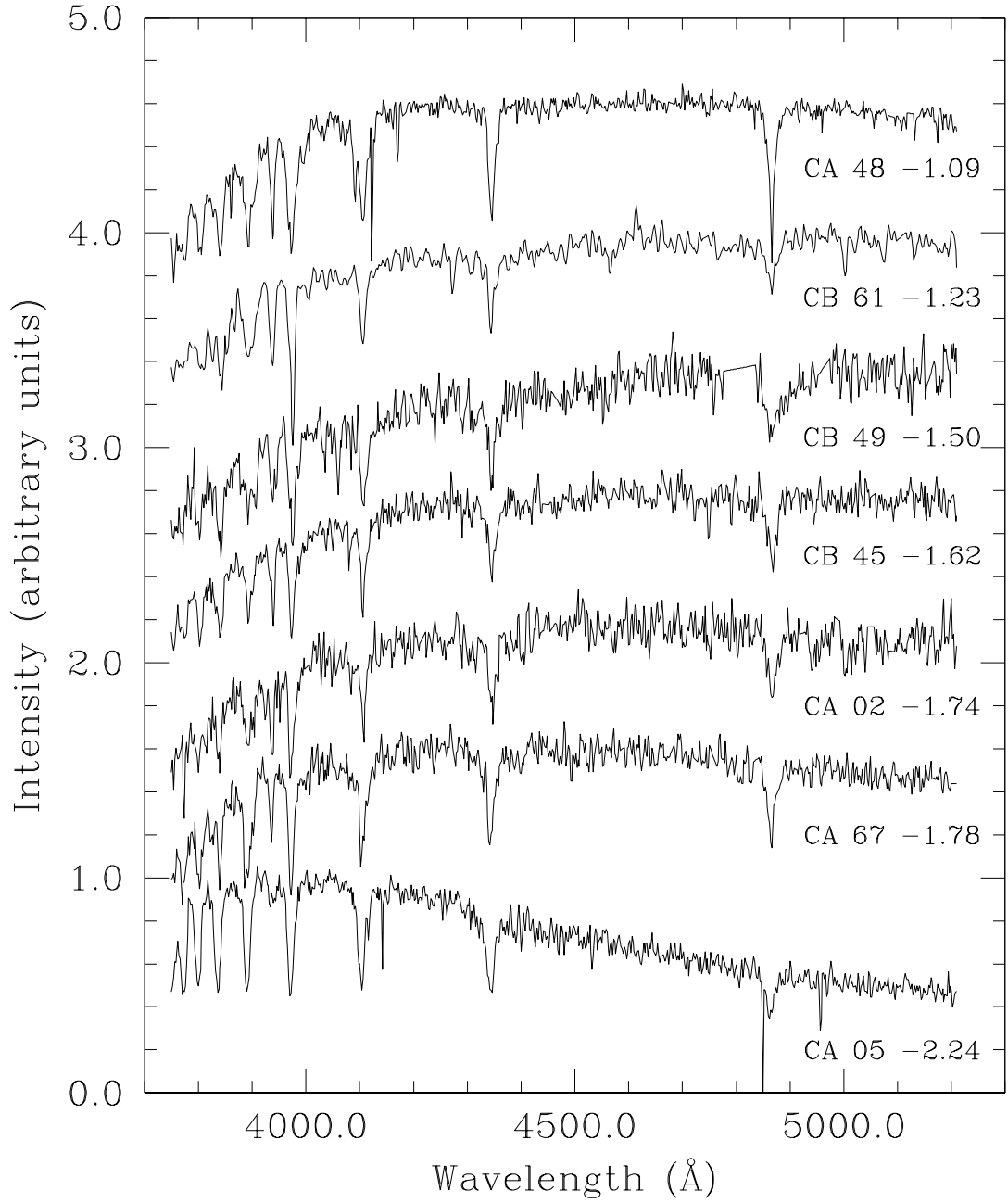}
\caption{Spectra of the 6 program stars on an arbitrary intensity
scale. They are shown in order of decreasing metallicity
from top to bottom.\label{fig-rrd}}
\end{figure}

Spectroscopic data were reduced using IRAF\footnote{ IRAF is distributed by
the NOAO, which are operated by AURA, under contract with NSF} and the
standard procedure for long slit spectra. Images were trimmed, bias
subtracted, and flat-fielded; spectra were traced, extracted and wavelength
calibrated. No flux-calibration was needed for our goals. We retained for
further analysis only the wavelength range 3750--5210 \AA, where
all lines of interest are located. The most delicate
parts of the whole procedure were spectrum extraction and background
subtraction, since the observed fields are very crowded and the seeing
conditions were not optimal. The risk of contamination from nearby objects is
high; this happened, for instance, in the case of one of our targets. The
background subtracted spectra of the 6 program stars are shown in 
Figure~\ref{fig-rrd}.

Spectral types were derived for our targets by comparing measured
pseudo-equivalent widths (EW's) of H${\beta}$, H${\gamma}$, and of the
\ion{Ca}{2} K line with those measured in the stars of Cr 140. The
pseudo-EW's were computed by dividing the instrumental fluxes within a small
spectral region ($f$) centered on the selected feature, with the average of
those in two comparison spectral regions ($c_1$ and $c_2$) located on both
sides of each feature and defining the local continuum (see
Table~\ref{tab-wl}). All spectra were shifted to zero velocity before
measuring the EW's. The (restframe) wavelength bins used are given in
Table~\ref{tab-wl}.

The spectral types for stars in Cr 140 given by CR are quite rough estimates.
To reduce this source of error they were averaged, whenever possible, with
the spectral types deduced from $uvby\beta$\ colors (Hauck \& Mermilliod
1998), using the calibrations by Crawford (1975, 1978, 1979). We then
constructed curves calibrating  EW's against spectral types using 13 of the
14 stars observed in Cr 140 (see Table~\ref{tab-cr140} for identification);
star \#17 was excluded because it is not on the main sequence. Finally, we
derived the best spectral type for each program spectrum by entering the
measured EW's for the program star and reading out the corresponding spectral
type. The final adopted H spectral type is the average of the values obtained
from H$\gamma$ and H$\beta$. Following Preston (1959), the observed \ds ~were
simply the difference between the H and Ca II K spectral types, in tenths of
spectral types. Results for the program stars are shown in
Table~\ref{tab-rrd2}, where we give for each of our targets the H and Ca
spectral types, and the \ds ~values derived applying both no phase corrections
(as suggested by Kemper 1982), or using phase corrections deduced from T Sex
(see Section 3.2).
The corresponding metallicities are also presented in Table~\ref{tab-rrd2},
and will be discussed in Sect. 4.

Our resolution is too coarse to separate the stellar and interstellar
components of the \ion{Ca}{2} lines, and we did not apply any correction for
the contribution due to interstellar absorption, since its effect on the \ds
~value, hence on [Fe/H], is negligible. Gratton, Tornamb\`e \& Ortolani (1986),
applying the \ds ~method to the globular cluster $\omega$~Cen, employed a
correction of 0.2 spectral subclasses (in the sense of an earlier spectral
type) to their \ds ~values, based on a reddening E(B--V)=0.11 mag. 
In our case the reddening is comparable (C2001), and the resulting
correction would imply a difference in the derived [Fe/H] of about 0.04 dex;
this is negligible with respect to the total error associated to the method.

To estimate scale and random errors in our derived \ds's, we have compared
the values we derived for the 6 comparison $ab-$type RR Lyrae's with those
given in the literature (see Table~\ref{tab-rrf} for individual values; the
comparison is done between our values and the average of literature \ds's
for each star). If we apply the phase corrections\footnote{
All phases in col. 2 of Table ~\ref{tab-rrf} were inferred from the Sp(H)'s
of our spectra, and agree with those computed from literature ephemerides in
all cases, but AF Vel and U Lep. These two stars have varying periods. Two
ephemerides are available for AF Vel and the corresponding phases (0.58: 
Eggen 1994, and 0.63: Lub 1979) are slightly later than we derive; we
suggest that both the ephemerides may be no longer adequate. Four ephemerides
are available for U Lep (0.60: Firmanyuk, Derevyagin \&  Lysova 1985;
0.67: Lub 1979; 0.69: Eggen 1994; and 0.80: Fernley, Skillen \& Burki
1983), and the phase we derive agrees with Eggen's prediction.
}
suggested by Butler (1975), our \ds ~are on average smaller than
the literature values by $0.2\pm 0.4$ tenths of spectral types, with an
$r.m.s.$ of 1.2 tenths of spectral type for individual stars. Since the
accuracy of the literature values is of the order of 0.5 tenths of spectral
types, we conclude that our \ds ~are on the standard scale, and have random
errors of $<1$\ tenth  of spectral type.

\subsection{Phase Corrections}

\begin{figure}
\vspace{10cm}
\includegraphics{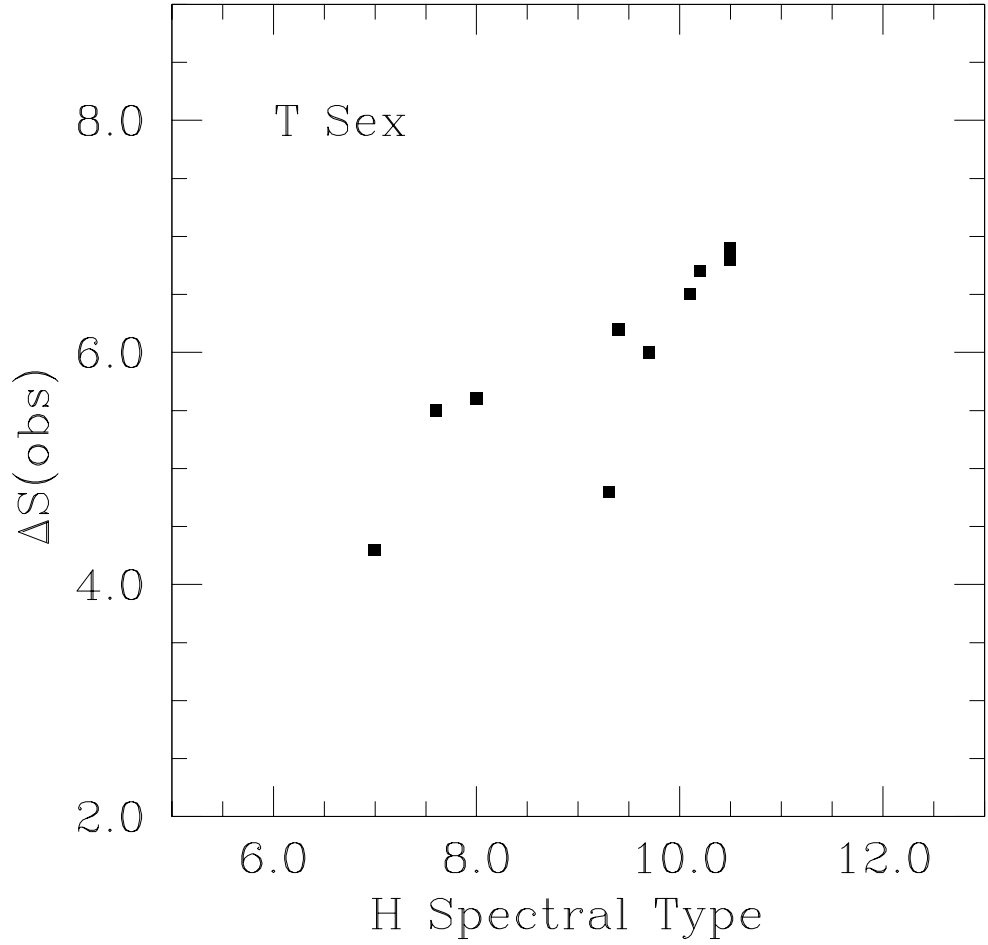}
\caption{T Sex: variation of SpT(H) and of the corresponding
\ds ~for the 10 spectra from which we have derived the phase corrections
described in Section 3.2.
\label{fig-tsex1}}
\end{figure}

Before interpreting the observed \ds ~in terms of metallicity, we have to
apply a correction for the observed phase, since \ds ~is known to vary during
the pulsation cycle. The exact form of the phase-correction to be applied for
$d-$type RR Lyrae's is not known. In their study of Galactic double pulsators
in the field, Clement, Kinman \& Suntzeff (1991) argued that the best
correction is the same used for $c-$type RR Lyrae's, since in general the
amplitude of the first harmonic is larger than the amplitude of the
fundamental mode for double pulsators (this is also true for our program
variables). Furthermore, the colors at minimum light of the $d-$type RR
Lyrae's are much closer to those of the $c-$type at minimum than to the
$ab-$type ones.
Unfortunately, the phase corrections to \ds~ for the $c-$type variables are
not very well defined. Kemper (1982) studied the variation of \ds ~along the
light curve of three $c-$type RR Lyrae's (RU Psc, T Sex, and DH Peg). He
concluded that phase corrections are small (of the same order of the
measuring error, i.e. $\pm$ 1 unit in \ds) and should not be applied when the
H spectral type is later than A7-A8, while he recommended to reject \ds
~estimates from spectra where the derived H spectral type is earlier. An
alternative procedure is to apply phase corrections  to the \ds ~according to
the phase variation of \ds ~derived from a well studied {\it c} type template
star. We have used to this purpose our observations of T Sex, and derive (see
Figure~\ref{fig-tsex1}):

\begin{displaymath}
\Delta S_{\rm cor} = \Delta S_{\rm obs}+0.587\,[10.5-SpT(H)],
\end{displaymath}

\noindent
where we give the correction simply as a linear function of the hydrogen
spectral type and, according to Kemper (1982), we neglect the small
difference existing between the ascending and descending part of the light
curve. This seems a reasonable assumption for the {\it c}-type pulsators (but
see Figure 1 of Smith 1986) given the fairly symmetrical shape of their light
curves, which all show slow climbs to maximum light, and because spectra of
RRc's do not seem to show the hydrogen line emissions and doubling which are
the signature of shock waves propagating through the atmosphere, found in the
spectra of the RRab's during the rise to maximum light (Preston \&
Paczyinski 1964, Chadid \& Gillet 1998, Gillet \& Crowe 1998).

Using the above relationship, we obtain for T Sex \ds=6.81$\pm 0.09$\, with an
$r.m.s.$\ scatter of 0.26 tenths of spectral types. This value is to be
compared with that determined using the procedure suggested by Kemper, that is
\ds=6.28$\pm 0.17$\, with an $r.m.s.$ scatter of 0.50 tenths of spectral
types. Also, for comparison the value given by Kemper for this star is $\Delta
S=6.1$\ (6.3 if the star were analyzed as an $ab-$type variable). However,
T~Sex has the largest variation of \ds ~with phase among the three stars
considered by Kemper, so this procedure may lead to somewhat too large phase
corrections (implying too large \ds's and too low metallicities). 
In Table~\ref{tab-rrd2} we give the \ds ~values of our targets obtained using
both the Kemper procedure (i.e. with no phase corrections and only using
spectra later than A7-A8) and the phase corrections derived from T Sex light
curve. On average, \ds ~values corrected according to the T Sex curve are
larger by $1.2\pm 0.2$\ spectral subtypes. Our adopted values are simply the
average of those determined using the two different procedures. We estimate
that uncertainties related to the phase correction are half the average
difference between these two estimates, i.e. about 0.6 spectral subtypes.

\section{Metallicities of the program stars}

The \ds ~values of our targets (both with and without phase corrections) were
translated to metallicities using two different relations. The first one
is given by Clementini et al. (1995, hereinafter C95), and is based on their
study at high resolution and high S/N spectra of 10 field RRab's, on data
adapted from Butler (1975) and Butler \& Deming (1979), and on globular
clusters having literature metallicities derived from high resolution
spectra:

\begin{equation} 
{\rm [Fe/H]} = -0.194\,\Delta {\rm S} - 0.08.
\end{equation} 

\noindent
The relation has been derived for $ab-$type RR Lyrae's, however Kemper (1982)
found that on average \ds ~values derived for $ab$\ and $c-$type variables in
GC's agree with each other, so that also $d-$type RR Lyrae's should
obey approximately the same \ds$-$[Fe/H] relation.
Metallicities derived from this relation are quite similar to the 
Zinn \& West (1984, hereafter
ZW84) metallicity scale for GC's.

The second one is given by  Gratton (1999, hereafter G99), and recalibrates the
\ds ~index using the new metallicity scale for GC's found by Carretta \&
Gratton (1997, hereafter CG97), which differs from ZW84's especially at
intermediate metallicities, usually giving somewhat higher metallicities. 
The metallicity dependence on \ds ~is in this case

\begin{equation} 
{\rm [Fe/H]} = -0.176\,\Delta {\rm S} - 0.03,
\end{equation} 
which is only marginally consistent with the relation in C95.
In the following we will be using metallicities derived from both relations:

(a) C95:  \ds ~values, both using no phase correction, as suggested by Kemper, and
correcting using T Sex, are given in columns 4 and 5 of Table~\ref{tab-rrd2},
and the adopted \ds ~(the average of the two) is given in  column 6. Columns 7
and 8 of Table~\ref{tab-rrd2} give the metallicities of the program stars
computed using the \ds ~estimates according to both Kemper procedure and T Sex
phase corrections, respectively, adopting C95 relation.
In the case of variable CA 48, we cannot give any metal abundance
using Kemper procedure for the spectrum taken near maximum light,
since the H-spectral type at which this star was observed is too early;
the lower S/N spectrum, instead, can be used to determine \ds ~values with
both methods. Both spectra give similar results, and in the following we
will be using the metallicity derived from the higher S/N spectrum.
Column 9 contains the adopted metallicities on C95 scale: they are simply the
average of the determinations obtained with the two different procedures. We
estimate that internal uncertainties in these metallicities are of about $\pm
0.2$~dex (from typical errors of $\pm 1$~in our \ds~ values). Systematic errors
are mainly due to uncertainties in the phase corrections ($\pm 0.1$~dex) and in
the metal abundance calibrations. The latter are however small (likely $\pm
0.1$~dex) in the abundance range of interest for the program stars. 

The average
metal abundance of the program stars is [Fe/H]=$-1.46\pm 0.09$\ (5 stars,
$r.m.s.$ = 0.21 dex) when the Kemper procedure is used. A somewhat lower average
value of [Fe/H]=$-1.60\pm 0.13$\ (6 stars, $r.m.s.$ = 0.33 dex) is obtained
when the phase corrections appropriate for T Sex are applied to the sample.
Note that in the latter case we include in the average one more star (variable
CA 48), which happens to be the most metal-rich of the sample. Finally, if we
use the adopted [Fe/H]'s shown in col. 9, we obtain an average [Fe/H]=$-1.49\pm 
0.11$\ (6 stars, $r.m.s.$ = 0.28 dex).
If we include the systematic errors present in our determinations, we conclude
that, on C95 scale,  our sample of 6 $d-$type RR Lyrae's in the bar of the LMC
has an average [Fe/H]=$-1.5\pm 0.2$\ .

This compares rather well with the results obtained by Alcock et al.
(1996). They used a similar method (line strengths of  \ion{Ca}{2} K and of
H$\delta$ measured on medium-low resolution spectra of quite low S/N) and
applied it to 15 field LMC RR Lyrae's. Alcock et al. do not expand much on the
subject: they do not give RR Lyr types, or individual values for metallicities.
They only say that the most frequent [Fe/H] value is about --1.6, with values
in the range from --2.4 to --0.8 (see their fig. 8), and that the estimated
accuracy is $\pm$0.25 dex.

(b) G99: Columns 10  and 11  of Table~\ref{tab-rrd2} contains the analogues of
cols. 7 and 8, but using G99 relation, and col. 12 shows the adopted [Fe/H]'s,
a simple  mean of the two above. 
In this case the average values are
[Fe/H]=$-1.29\pm 0.07$\ (5 stars, $r.m.s.$ = 0.16 dex) adopting Kemper
procedure,
[Fe/H]=$-1.41\pm 0.12$\ (6 stars, $r.m.s.$ = 0.30 dex)
adopting phase corrections derived from T Sex, and
[Fe/H]=$-1.32\pm 0.10$\ (6 stars, $r.m.s.$ = 0.25 dex)
adopting the recommended values in col. 12.
Again including systematic errors, we then conclude that, on G99 scale,  our
sample has a somewhat larger average metallicity, of [Fe/H]=$-1.3\pm 0.2$\ .

The value compares slightly worse to Alcock et al. (1996), but this is not
surprising, since CG97 scale tends to produce higher metallicities at this 
intermediate metal abundance. Anyway,
15 and 6 objects are too small samples to attach significance to a $\sim$
0.3 dex difference, well within the quoted uncertainties.

Notice that we have two RRd's with [Fe/H] $>-1.5$ (C95) or $>-1.3$ (G99), hence
with metal abundances larger than the Oo I clusters M 3 and IC 4499: these are
the most metal-rich RRd's identified so far.

\section{Masses}

\subsection{The Petersen diagram}

Masses for the double mode pulsators can be evaluated from the ratio between
first overtone (P$_1$) and fundamental (P$_0$) pulsation periods. Petersen
(1973) introduced the use of what is now universally known as ''the Petersen
diagram'', where the ratio P$_1$/P$_0$ is plotted versus the value of P$_0$ for
each RRd.
Pulsation models define loci of constant mass in this diagram, hence RRd masses
can be determined by the position of the star in the Petersen diagram,
interpolating/extrapolating between these models.

Figure~\ref{fig-p0p1} shows the position in the Petersen diagram of the six LMC
RRd's studied in this paper (we have omitted from the figure the remaining 67
LMC RRd's in A97 sample for clarity, but our small sample is well representative
of the general P$_1$/P$_0$  versus P$_0$ distribution of the LMC RRd's, see
Figure~\ref{fig-p0p1zoom}), together with the other RRd's found in our Galaxy
(clusters and field) and in the Draco and Sculptor dwarf spheroidal galaxies.
Data plotted in Figure~\ref{fig-p0p1} have been taken 
from Garcia-Melendo \& Clement (1997) for NSV09295; 
from Clementini et al. (2000) for CU Com;
from Clement et al. (1991, 1993) for AQ Leo, VIII-10, and VIII-58; 
from Clement et al. (1993) for NGC 2419 (1 object), and NGC 6426 (1 object); 
from Corwin, Carney \& Allen (1999) for M 3 (5 objects); 
from Walker \& Nemec (1996) for IC 4499 (17 RRd's);
from Walker (1994) for M 68 (12 RRd's); 
from Nemec (1985b) for M 15 (14 RRd's); 
from Nemec (1985a) for Draco (10 RRd's); 
from Kaluzny et al. (1995) for Sculptor (1 object);
from A97 for the LMC (73 RRd's).

This figure is the analog of fig. 2 in A97; they also plotted the latest 
pulsational models available in literature, by Bono et al. (1996, henceforth
BCCM96) computed for metallicity Z=0.0001, and for the three masses 0.65, 0.75
and 0.80 \msun.
BCCM96 computed non-linear, non-local, time dependent pulsational models based
on up-to-date opacities (Rogers \& Iglesias 1992). These models, whose
properties and physical and numerical assumptions are described in Bono \&
Stellingwerf (1994), in BCCM96 and in Bono et al. (1997a,b) are claimed by 
authors to reconcile values for masses and luminosities based on pulsation and
on stellar evolution theories.

However, as already noted by A97, several objects (about 40 of the 73 double
mode LMC RR Lyrae's, among which our CA 02, and one of the RRd in M 3) fall in
the region of the Petersen diagram below the 0.65 M$_\odot$ BCCM96 model, and
extrapolation of the BCCM96 models would produce rather small masses of about
0.55 \msun ~for these objects.
This would not be simple to explain for the stars in this region with measured
metallicity: M3 is the prototype Oosterhoff I cluster, and CA 02 and the
Galactic field star VIII-58 have similar abundances, intermediate between Oo I
and Oo II clusters ([Fe/H] $\simeq$ --1.7 on C95 scale, or --1.5 on G99 scale).
A97 discuss somewhat the problem, also suggesting, but in the end discarding,
the possibility of these stars having a larger metal content, hence a mass
larger than obtained by the BCCM96 models.

\subsection{The pulsation models: derivation of the masses}

\begin{figure}
\vspace{15cm}
\includegraphics{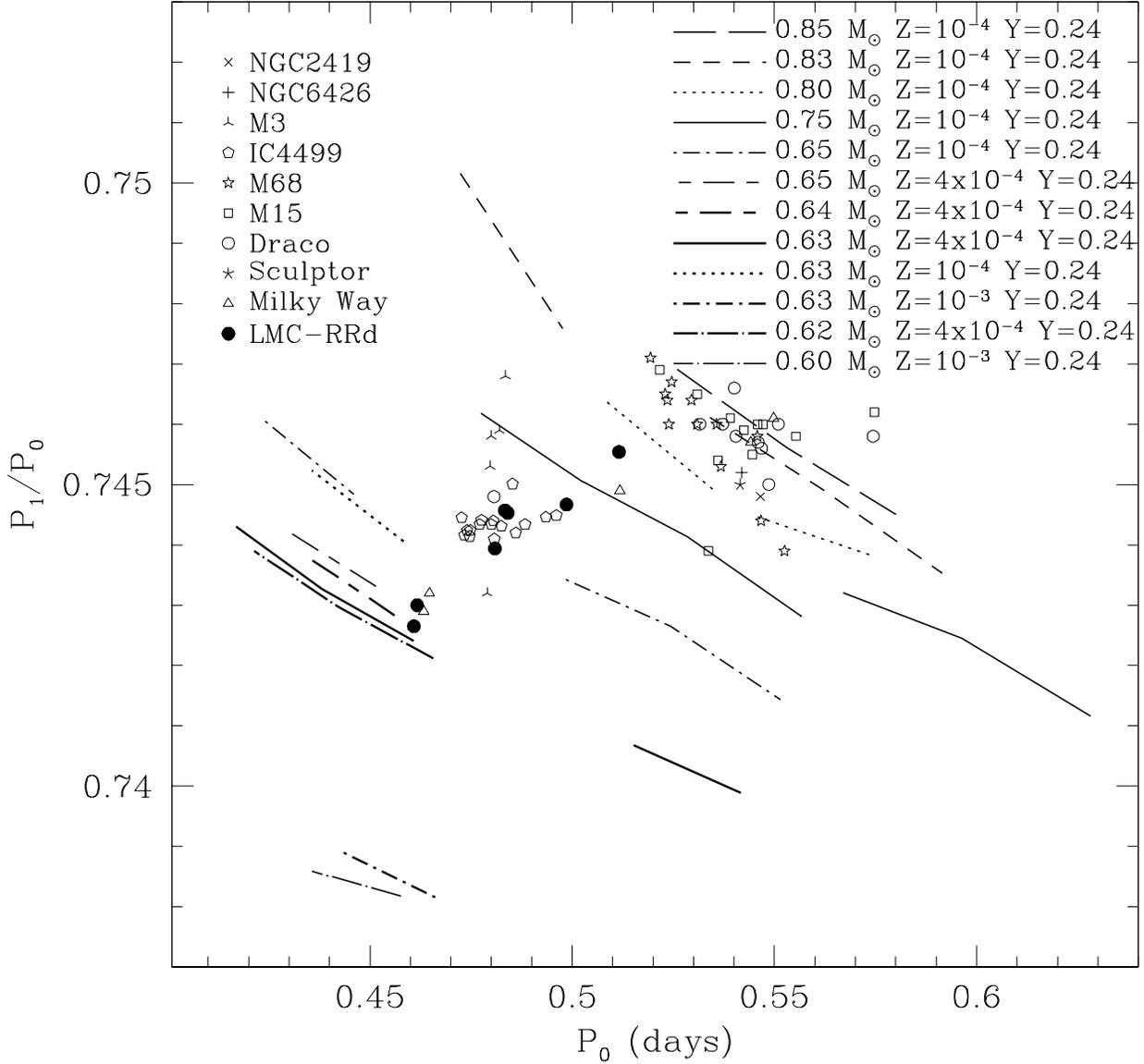}
\caption{The Petersen diagram for known RRd's in our Galaxy
(both in the field and in globular clusters), in the two dwarf spheroidals
Draco and Sculptor, and in
the LMC, indicated by different symbols. Also plotted are the BCCM96 models (Z
= 0.0001, and masses 0.80, 0.75, and 0.65 M$_\odot$) and some of the new
models  extending them to different metallicities and masses (See Section 5
for an explanation).
\label{fig-p0p1}}
\end{figure}

\begin{figure}
\vspace{15cm}
\includegraphics{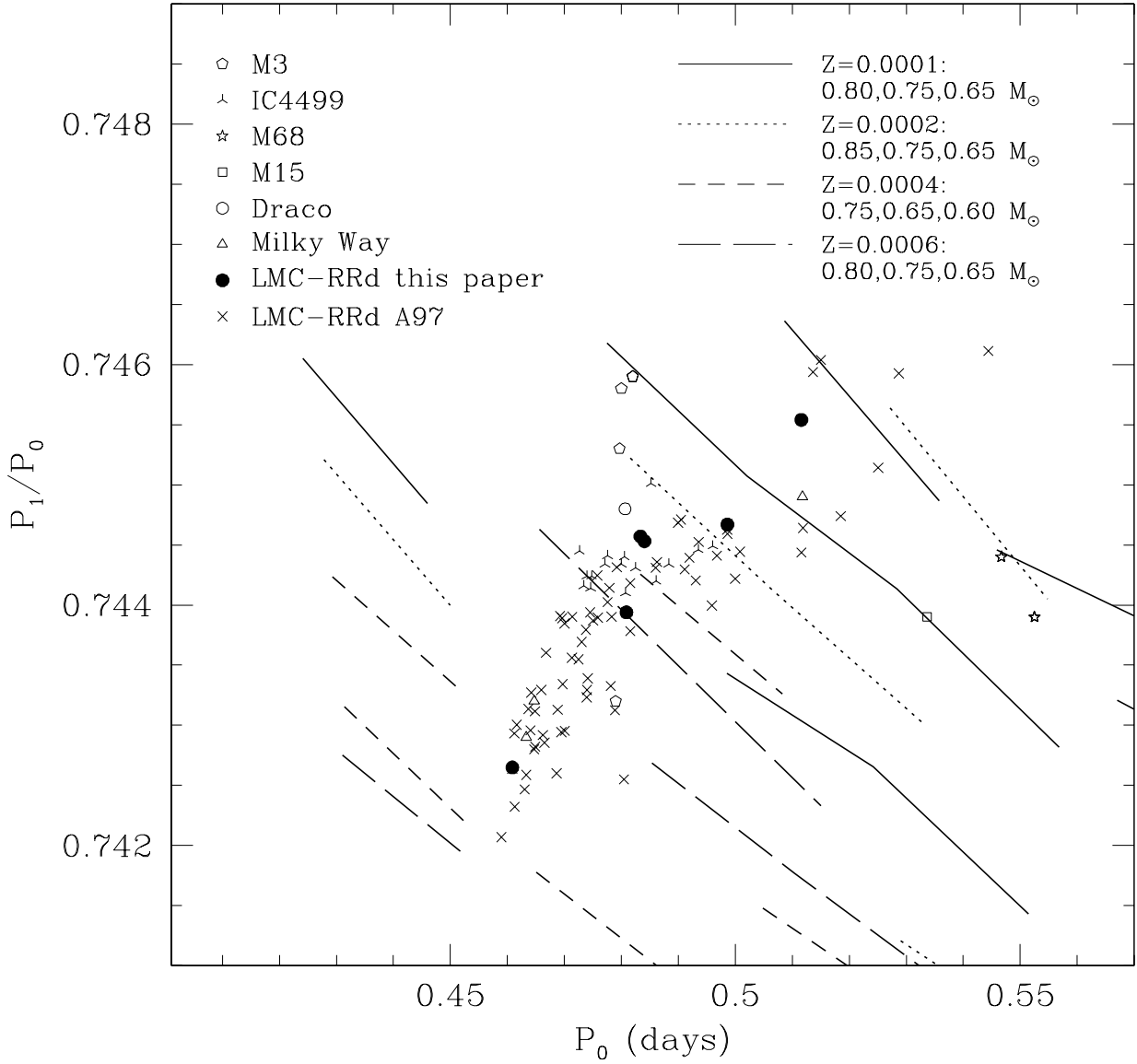}
\caption{Enlargement of Figure~\ref{fig-p0p1} in the region
relevant for LMC RRd's: here all LMC RRd's found by A97 are plotted.
\label{fig-p0p1zoom}}
\end{figure}

Since we now have information on the metallicity of 6 LMC RRd's, we have
decided to test the dependence of masses derived from the Petersen diagram 
also on metallicity. In order to do this, we  derive masses for all objects in
Figure~\ref{fig-p0p1} for which the metallicity has been estimated, using the
most appropriate models. 
Metallicities for all objects were adopted as follows:
in the case of the LMC RRd's we used the \ds ~values (this paper), as for
three MW field RRd's (AQ Leo, VIII-10, VIII-58: Clement et al. 1991), and
converted them to [Fe/H] using both C95 and G99 relations; 
in the case of CU Com we used the [Fe/H] value in Clementini
et al. (2000), obtained from high resolution spectroscopy (note that at these
low metal abundances ZW84 and CG97 scales coincide); 
for the dSph's Draco and Sculptor we took the literature values (Mateo 1998)
on ZW84 scale, and also converted them to CG97 scale according to the 
transformation relations provided by GC97 (see eq. 7 in that paper);
for Galactic GC's, since \ds ~values were not available for all of them, we
used the [Fe/H] values  available in the literature (ZW84 for all of them,
CG97 direct measurements for M 3, M 15, M 68, and transformation to CG97 scale
for NGC 2419, NGC 6426, and IC 4499).
Metallicities chosen in the above way should be on a homogenous base for
field and cluster variables, 
since the metallicity derived from the C95 relation
(equation 1) is close to the ZW84 scale, while that derived from the G99
relation (equation 2)  is close to the CG97 one.
%C95 relation between \ds ~ and metallicity
%is more closely on ZW84 scale, and G99  on CG97 one.

Since both mass and metallicity affect the predicted Petersen diagram, 
information about the 
metallicity  is required for a proper evaluation of the
``pulsational'' mass. At the same time, for each derived stellar mass the
luminosity level can be inferred on the basis of the fundamental  period
range.  We note that the only pulsation models that populate the Petersen diagram
are those located in the double mode (OR) region of the predicted instability strip, 
that is betwee
the theoretical fundamental blue edge and the theoretical first overtone red
edge, for each assumed luminosity level.
For each mass, models with different luminosity levels were computed in order
to cover the full range of evolutionary predictions (which are dependent on the
adopted input physics) for the luminosity of horizontal branch stars and also
to take into account evolutionary effects (RR Lyrae evolving from their zero
age horizontal branch location). As already noticed by BCCM96 nonlinear
computations offer the opportunity to disentangle the mass and luminosity
effects on the location in the Petersen diagram. As shown in 
Figure~\ref{fig-p0p1}, once the mass and metallicity are fixed, the period
values identify the luminosity level.
However we consider the luminosity derivation beyond the aims of the
present paper.

Models with Z=0.0001, 0.0004, 0.0006, 0.0008 have been computed 
for different masses and luminosities, using the same treatment as in BCCM96.
In particular, we have used the following models:
i) Z = 0.0001 at 0.85, 0.83, 0.80, 0.75, 0.65, 0.63 \msun ~and with 3 luminosity
levels, log L = 1.61, 1.72, 1.81; 
ii) Z = 0.0004 at 0.75, 0.70, 0.65, 0.64, 0.63, 0.62, 0.60 \msun ~and 3 
luminosity levels, log L = 1.61, 1.65, 1.72;
iii) Z = 0.0006 at 0.80. 0.75, 0.65 \msun ~and two luminosities, log L = 1.61, 
1.72; 
iv) Z = 0.0008 at 0.85, 0.65 \msun ~and 3 luminosities, log L = 1.61, 1.72,
1.81.
A few additional models at Z = 0.0002, 0.001, and 0.002 have also been
computed, but not extensively used to determine masses. The main parameters
of the adopted models (Z, mass, luminosity, P$_0$, P$_1$) are given in
Table~\ref{tab-mod}, available only in electronic form.

A subset (indicated by a star symbol in Table~\ref{tab-mod}) 
of the Z=0.0001, 0.0004, 0.0006 and 0.0008 models is shown in
Figure~\ref{fig-p0p1} where, for each metallicity, the upper curve refers to
the higher mass, and the  luminosity levels increase from right to left; now
the models completely cover the range observed for the RRd's.
An enlargement of the region of interest for our 6 LMC RRd's is 
shown in Figure~\ref{fig-p0p1zoom}, where all the LMC RRd's from A97 are 
also plotted.

We derived masses for our targets as well as for all RRd's for which a metal
abundance estimate is available (either direct or of the hosting system)
interpolating between the set of pulsational models at the four indicated
metallicities. For each object we obtained a table with four mass values, one
for each Z, and a fit with a parabola produced a quadratic relation mass $vs$
Z.
We then entered the empirical [Fe/H] value for the star (we do it for both
metallicity scales used in this paper) and derived the mass with the
associated error (based on the error on the fit and an assumed 0.2 dex error on
[Fe/H]).
In some cases this procedure was slightly modified because the parabolic
interpolation was not completely stable: for NGC 2419 and NGC 6426 we also
used the Z=0.0002 models; for CU Com, the most metal-poor object of the
sample, we found more adequate to use only the models at Z=0.0001, and
similarly for CA 48 and CB 61 we used only the models at Z=0.0008, and for CB
49 only the models either at Z=0.0006 (on C95 scale) or at Z=0.0008 (on G99
scale).
Results for masses and associated errors for the 6 LMC RRd's, for the MW field
RRd's, and for the two dSph's are given in Table~\ref{tab-rrd3}, while
Table~\ref{tab-rrd4} presents results for galactic clusters RRd's.

Note that mass values for M 3 are the most uncertain because of the lower
precision of the periods P$_0$ and P$_1$ given for its RRd's.
However, this could not
be an explanation for the rather unusual masses found for the two Oo I
clusters, since IC 4499 has exactly the same behaviour, but periods as precise
as those of the other GC's.

\subsection{The mass-metallicity distribution}

\begin{figure}
\vspace{15cm}
\includegraphics{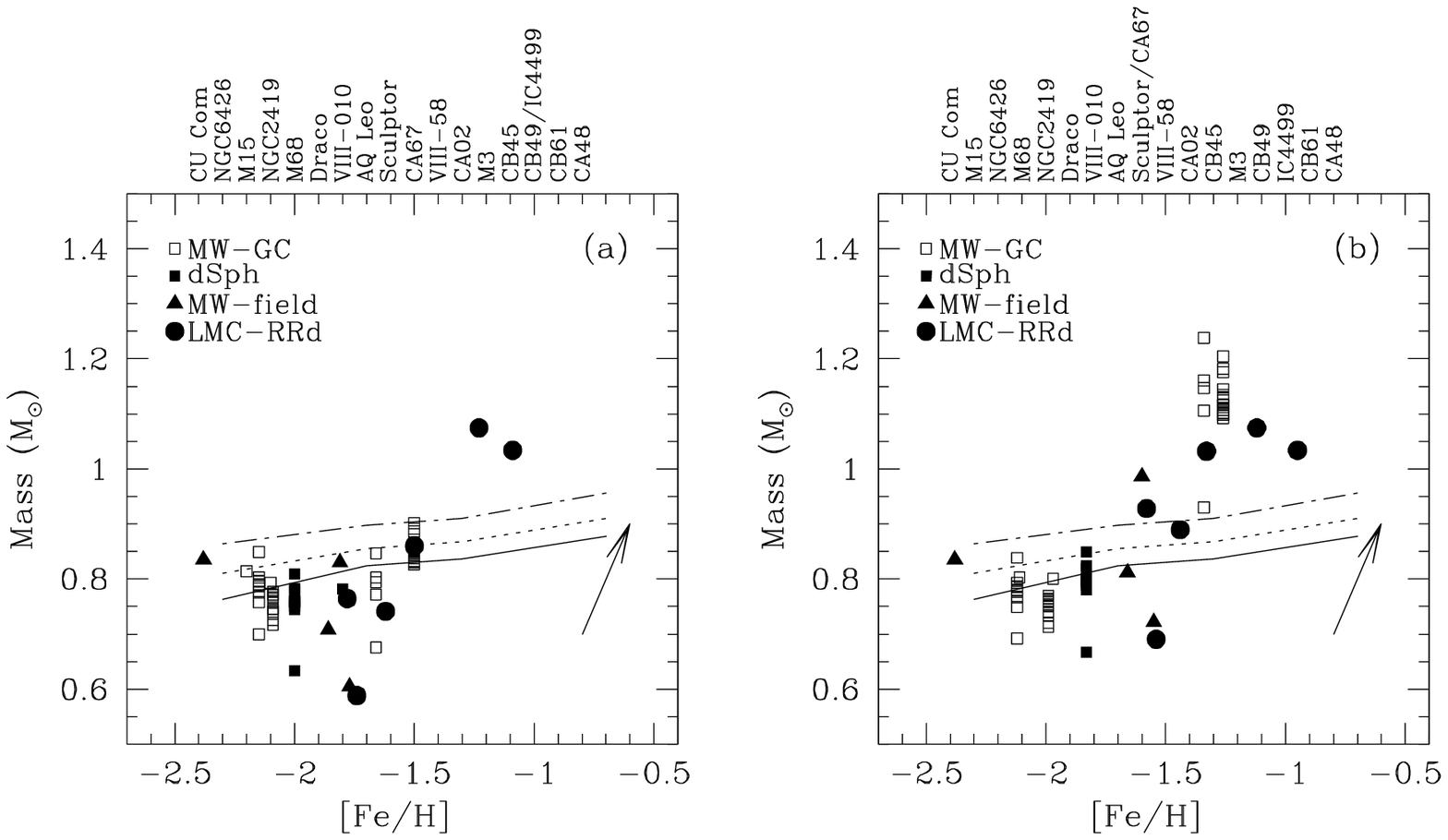}
\caption{Mass - metallicity plots: (a) using
metallicities on the ZW84 scale for GC's and dSph's, and C95 relation for RR
Lyrae's; (b) using metallicities on CG97 scale for GC's and dSph's, and G99
relation for RR Lyrae's.  See discussion in Section 5.3 for extensive
comments on the high mass of the Oo I clusters.
In both panels the program stars are indicated by filled circles, field RRd's
in our Galaxy by filled triangles, RRd's in GC's by open squares, and dSph's
by filled squares. References for the original papers are given in Section 5.
Lines indicate the variation of mass at the RGB tip with metallicity, for ages
of 14 (solid line), 12 (dotted line), 10 (dot-dash line), 8 (long dash line),
and 6 (dot-long dash line) Gyr, as deduced
from Bertelli et al. (1994) and Girardi et al. (1996).
The arrows indicate the effect on mass determination of a 0.2 dex error in
the assumed [Fe/H].
%nb uses fignewmasmet3_r1.ps
\label{fig-masmet} }
\end{figure}

Figure~\ref{fig-masmet} shows graphically these results; in panel (a)
[Fe/H]'s are on C95 scale, and in panel (b) on G99 metallicity scale. 
The almost flat distribution of masses with varying metallicities,
particularly in Figure~\ref{fig-masmet}(a), shows no indication of 
a mass - metallicity
relation. In fact, mass and metallicity have an opposite effect on the
position of a RRd in the Petersen diagram, leading to a sort of degeneracy.
A mass - metallicity relation is only obtained using a fixed metallicity
to compute the pulsational models, then using the derived masses in
connection with the empirical abundances.

This is what was often done in the past (e.g., BCCM96, and fig. 1 in Cox
1991), but it implies neglecting the effect of the metal abundance on the
position in the Petersen diagram, hence an error on the derived mass for all
objects whose metallicities differ from the metal abundance adopted in the
pulsational model computations. Petersen (1991), on the basis of linear
calculations, found that the difference in metal abundance between Oo I and
Oo II clusters produces a very small mass variation, whereas Cox (1991)
predicts that the mass of all globular cluster RRd variables should be close
to 0.8  \msun ~when metallicity is taken into account and the  OPAL opacity
tables for the proper composition are used. Our nonlinear computations show
that the metallicity effect on the Petersen diagram is strong, since it acts
in different ways on the fundamental and first overtone periods, and heavily
influences their ratio. This result is in agreement with the predictions by
Kovacs, Buchler \& Marom (1991) and Cox (1995), who also found a noticeable
dependence of the Petersen diagram on metallicity when the new OPAL opacities
are used in pulsation computations. In particular Cox (1995) performs
computations with Z=0.0003 and find a mass increase of OoI RRd with respect
to those derived from models at Z=0.0001. However he reproduces the IC 4999
RRd's location with Z=0.0003, M=0.70 $\pm$ 0.05 \msun ~models, whereas the
metallicity used in our paper for IC 4999 RRd's is equal or larger than
0.0006, depending on the metallicity scale. This most likely justifies the
larger masses we find for OoI RRd's with respect to Cox (1995) results. In
fact, as shown in Figure~\ref{fig-p0p1zoom}, our Z=0.0002-0.0004, M = 0.75
\msun ~models fit the IC 4999 RRd's location too.

Further analysis on the metallicity dependence of pulsational models is
required, since there is still the possibility that models could somehow
overestimate the effect as a result of the adopted opacity tables (the most
recent Livermore ones, Iglesias \& Rogers 1996; see also the discussion by
Cox 1991) or that, as suggested by Kovacs et al. (1992)  the Petersen diagram
depends not only on the heavy element abundance, but also on the chemical
mixture, but this issue is quite beyond the scope of the present paper.

In Figure~\ref{fig-masmet}(a), the globular clusters of both Oosterhoff groups
have similar masses, with very similar internal scatter, while in
Figure~\ref{fig-masmet}(b) the two Oo I clusters M3 and IC 4499 show a
(unrealistically?) higher mass than the Oo II GC's, as a result of the
different (higher) metallicity scale adopted. 
Indeed, some weak hints of a mass metallicity relation with larger masses at 
higher metal abundances could be seen also among the Oo I field and cluster 
Rd's in Figure~\ref{fig-masmet}(a).
Whether this finding might imply an actual
difference between GC's, or is simply to impute to a too high sensitivity of
the model-derived masses on metallicity, not well tuned yet, is an open
question. 
The offset between the two metallicity scales is about 0.2 dex at the typical
metallicity of the Oo I clusters; comparison between Figure~\ref{fig-masmet}(a)
and Figure~\ref{fig-masmet}(b) tells quite clearly that a systematic error of
0.2 dex in the metallicity of the Oo I clusters, through the effect on the
RRd's masses  derived from pulsational models, goes in the direction of
producing a mass-metallicity relation (of any kind) that might actually not be
there.
The arrows in Figure~\ref{fig-masmet} show graphically the 
variation of derived masses for a 0.2 dex increase in the assumed [Fe/H],
comparable to the error associated to abundances obtained with the \ds
~method.

The derived masses are not in contrast to values expected from stellar
evolutionary models for an age of 10-14 Gyr, and no extensive mass
loss on the red giant branch. This is shown in Figure~\ref{fig-masmet},
where the lines represent the mass at the RGB tip for ages of 6, 8, 
10, 12, and 14 Gyr; they are taken from the evolutionary models by Bertelli et
al. (1994) for Z=0.004, 0.001, 0.0004, and by Girardi et al. (1996) for
Z=0.0001, however any other choice of isochrone sets would produce similar
results.
Within the errorbars (see Tables~\ref{tab-rrd3}, and ~\ref{tab-rrd4}) all
objects in Figure~\ref{fig-masmet}(a) are well consistent with the plotted
isochrone sets. Only the two most metal rich RRd's of the LMC lie above the
10 Gyr isochrone; however, the large errorbar associated to their mass 
estimates on one hand, and/or the possibility for these field objects to 
be slightly younger that 10 Gyr might account for their position in the 
figure. Figure~\ref{fig-masmet}(b) is more difficult to explain, since
RR'd in both M3 and IC4499 lie well above the 10 Gyr isochrone, even 
taking into account errorbars,
and ages younger than 10 Gyr might be at the lower limit of the acceptable 
values for these GCs.

This could be due to uncertainty in the pulsational masses:  models could
have a too high sensitivity to abundances at these
relatively large metallicities, either through the total heavy element
abundance or the adopted chemical mixture. 
Another possibility is a too large abundance assumed for
these clusters  (e.g., an overestimate of about 0.15 dex in CG97 
computations for intermediate-metallicity clusters).
This last possibility can be checked for M 3, which has also been studied
with high resolution spectroscopy and fine abundance analysis by Kraft et
al. (1992); they find [Fe/H]=$-1.47\pm 0.01$, to be compared to
[Fe/H]=$-1.34\pm 0.02$ in CG97. But also adopting the Kraft et al. value, the
pulsational masses are larger than for the Oo II clusters. Furthermore, work on
high resolution spectra of Turn Off stars in NGC 6752, obtained with UVES on
the VLT (Gratton et al. 2001), obtains the same iron abundance already derived
for giants in CG97 for this intermediate metallicity cluster ([Fe/H]=$-1.42$).
Small overestimates may be present, but do not seem to
explain completely the effect noted in Figure~\ref{fig-masmet}(b). 
Finally, we wish to note that, based on a recent study of M 92
by King, Stephens \& Boesgaard (1998), and on spectra of Turn Off stars in NGC
6397 (Gratton et al. 2001), it may be necessary to reassess also the low
metallicity end of the GC scale, in the sense of lower abundances: this would
go in the direction of increasing the metallicity difference between Oo I and
Oo II clusters.
Discussing the validity of metallcity scales is outside the scope of the
present paper, and we prefer to derive masses using different ones,
leaving to the reader the final choice about which one to believe better.

In any case, the major result here is that masses of Oo I and Oo II clusters
do not significantly differ, or, if they do, they differ in the sense of Oo I
GC's  having higher masses, contrary to what usually accepted. This stems from
the use of appropriate metallicities in the pulsational models. This result is
summarized in Table~\ref{tab-rrd5} where we also compare the average mass values
derived for the field and cluster variables of differing Oosterhoff types,
adopting the two metallicity scales, separately. The cut between
Oosterhoff types was set at [Fe/H]=$-$1.7 and $-$1.5 in ZW84 and CG97
respectively, with the Oo II variables being at [Fe/H]$\leq -$1.7 ($-$1.5).

The field RRd's have a larger scatter, both in metallicity and in mass;
however most stars lie in the same region of the diagram occupied by the
globular cluster stars. The bulk of the field Galactic RRd's as well as RRd's
in Draco and Sculptor concentrate in the low metallicity region occupied by
the Oo II clusters, while almost all of the LMC RRd's in our sample are in the
region of the Oo I clusters, or extend further towards  metallicity values
which have no counterparts in our Galaxy or in the two dSph's where RRd's had
been found in the past.
This confirms C2001 finding that pulsational properties of the RR Lyrae
variables in their observed regions of the LMC bar seem to follow the
period-amplitude relation of Oo I clusters like M 3.

Finally note that, no matter which metallicity scale is adopted, cluster and
field stars of similar metallicity and/or Oosterhoff type do not show any 
systematic difference in the derived mass.

\section{Summary and conclusions}

In order to investigate the possibility of a systematic difference in the mass
and mass-metallicity distribution for RR Lyrae in globular clusters and in the
general field, we have used the Preston \ds ~method to derive metallicities for
6 RRd's (double pulsating variables) in the bar of the LMC. We have then
combined these values with literature data for four field Galactic RRd's,
making up a total of 10 field RRd's whose metallicity has been directly
measured, and with data for RRd's in the Draco and Sculptor galaxies, for which
the same metallicity as the host galaxy has been assumed. For these stars,
pulsational masses were derived using an extension of the BCCM96 models to
enlarged mass and metallicity ranges, purposefully computed for this paper.
The same procedure has been applied to globular clusters RRd's, both of Oo I
and Oo II types, similarly finding their masses.

We have then compared the position of field and cluster RRd's in the
mass-metallicity diagram: we find that, on average, the two samples follow
the same mass-metallicity distribution.

Since field and cluster RR Lyrae's also obey the same
mass-luminosity-metallicity relation (Catelan 1998, Carretta et al. 2000a), we
conclude that they should also obey the same luminosity-metallicity relation,
and that there is no difference in luminosity between field and cluster RR
Lyrae's. This result implies that results found for field RR Lyrae can be 
safely used also to derive properties of cluster RR Lyrae's, like e.g. the
absolute luminosity and the absolute luminosity-metallicity relation.

\acknowledgments{
We thank G. Bono for having provided data in tabular form 
and M. Tosi for very useful 
discussions on the mass-metallicity relation between Oosterhoff groups. 
We also wish to thank the anonymous referee for the
suggested improvements which clarify our meaning.
This research has made use of the Simbad database, operated at CDS, 
Strasbourg, France. This work was partially supported by MURST - Cofin98
under the project "Stellar Evolution."
}

\newpage

\begin{table}[t]
\caption{Information on the observed RRd's
\label{tab-rrd1}}
{\scriptsize
\vspace*{5mm}
\begin{tabular}{lccclccccl}
\tableline
\tableline
Name  &Macho ID &$\alpha$&$\delta$& $<V>$ & P$_0$ &P$_1/P_0$ & 
Epoch&HJD&~~$\varphi$\\
&&2000&2000&mag&day&&$-$2451100&$-$2451100&\\
&&&&&&&day&day&\\
\tableline
CA02 (23032) & 13.6691.4052  & 5 21 34.1 & $-$70 31 52.1 &19.67& 0.46087 & 
0.74266 &83.62317&97.79014&0.74\\       
CB45 (7467)  & 13.6080.591   & 5 18 16.6 & $-$70 55 17.6 &19.01& 0.48089 & 
0.74394 &83.81805&97.62984&0.72\\      
CA48 (4420)  & 06.6811.651   & 5 22 45.2 & $-$70 36 35.7 &19.35& 0.48336 & 
0.74457 &83.70869&97.74300&0.03
\tablenotemark{*}\\      
      &               &        &               &     &        &         &        
   &96.55699&0.58\\      
CB49 (3347)  & 13.5836.525   & 5 16 27.0 & $-$71 02 32.8 &19.19& 0.48406 & 
0.74453 &85.67901&97.54534&0.51\\        
CB61 (4509)  & 13.5958.518   & 5 17 37.5 & $-$71 00 26.3 &19.49& 0.49861 & 
0.74467 &84.73568&97.58639&0.77\\        
CA67 (3155)  & 06.6810.428   & 5 22 35.9 & $-$70 38 28.4 &19.14& 0.51159 & 
0.74555 &86.70999&96.73102&0.59\\        
\tableline                                                                  
\normalsize
\end{tabular}
\tablenotetext{*}{Two spectra are available for CA 48. The one at
HJD=51197.74300 was observed around maximum light, while that taken at
HJD=51196.55699 ($\varphi$=0.58) is of low S/N. However, the phase corrected \ds
~we derive from the spectrum at early phase (5.2) is in extremely good
agreement with that derived from the low S/N spectrum (5.3; see column 6 of
Table~\ref{tab-rrd2})}
}
\end{table}

\begin{table}[ht]
\caption{Informations on the spectral type standard stars observed in Cr 140 
\label{tab-cr140}} 
\vspace*{5mm}
{\small
\begin{tabular}{rcrcccc}
\tableline
\tableline
No. &   V  &  B$-$V    &  $b-y$ & Sp.Type & Sp.Type & Sp.Type \\
    &      &         &        & (CR)   &($b-y)$   & (adopted) \\
\tableline
 6   & 7.12 & --0.04 &         & A0     &       & A0 \\
 8   & 7.49 &   0.02 &         & A2     &       & A2 \\
 9   & 7.59 & --0.09 & --0.042 & A0     & B6.3  & B8 \\
 15  & 8.59 & --0.06 & --0.020 & A0     & B9.3  & A0 \\
 16  & 8.68 &   0.40 &         & F2     &       & F2 \\
 17\tablenotemark{1}  & 8.70 &   0.70 &   0.467 & F5     & G0.2  &    \\
 19  & 8.83 & --0.02 & --0.017 & A0     & B9.7  & A0 \\
 22  & 8.97 &   0.42 &         & F2     &       & F2 \\
 24  & 9.00 &   0.19 &         & A3     &       & A3 \\
 26  & 9.15 &   0.20 &  0.111  & A2     & A5.5  & A4 \\
 29  & 9.28 &   0.14 &         & A2     &       & A2 \\
 32  & 9.34 &   0.02 & --0.005 & A0     & A9.9  & A0 \\
 38  & 9.57 &   0.22 &   0.140 & A5     & A7.5  & A6 \\
 42  & 9.71 & --0.07 &         & A0     &       & A0 \\
\tableline
\end{tabular}
}
\tablenotetext{1}{This star was observed, but not used, since it is not on
the main sequence}
\end{table}

\begin{table}[t]
\vspace*{7cm}
\caption{Spectral regions used to compute the pseudo-EW's
\label{tab-wl}}
\vspace*{5mm}
\begin{tabular}{lccc}
\tableline
\tableline
Line       & Blue Continuum      & Line Region         & Red Continuum \\
\tableline
\ion{Ca}{2} K & $c_1$=3850-3870~\AA & $f$=3925-3950~\AA & $c_2$=3940-3960~\AA\\
H$\gamma$     & $c_1$=4220-4280~\AA & $f$=4320-4360~\AA & $c_2$=4420-4480~\AA\\
H$\beta$      & $c_1$=4680-4780~\AA & $f$=4840-4890~\AA & $c_2$=4950-5050~\AA\\
\tableline
\end{tabular}
\end{table}

\noindent
\begin{table}[t]
\caption{Field {\it ab}-type RR Lyrae: comparison between literature and 
the newly derived \ds ~values
\label{tab-rrf}}
\vspace*{5mm}
\begin{tabular}{lccccccc}
\tableline
\tableline
Star   &$\varphi$ &\ds    &Reference &SpT(H) &SpT(K) &\ds   &\ds \\
       &          &liter. &liter.    &       &       & obs. & corr.  \\
\tableline
IU Car &  0.80  &    9      & (a)     &  F3.2 &  A5.2 &    8.0     &    8.2  \\
X Crt  &  0.77  &   10,10.4 & (a,d)   &  F3.2 &  A4.3 &    8.9     &    9.1  \\
       &  0.83  &   10,10.4 & (a,d)   &  F3.5 &  A4.7 &    8.8     &    8.8  \\
WY Ant &  0.68  &    6      & (a)     &  F2.2 &  A4.8 &    7.4     &    8.2  \\
AF Vel &  0.55(0.58-0.63)& 7      & (a)     &  F0.9 &  A5.2 &    5.7     &    
7.2  \\
U Lep  &  0.69(0.60-0.80)&    8,9,9.4,9.22 & (a,b,c,e) &  F2.6 &  A4.3 &    8.3  
   &    8.8  \\
TV Leo &  0.69  &   10,10,9.9,10.49,10.8  & (a,b,c,e,f)   &  F1.7 &  A3.1 &    
8.6     &    9.7  \\
\tableline
\end{tabular}
\tablerefs{(a) Lub (1979); (b) Preston (1959); (c) Butler (1975); (d) Kinman
\& Carretta (1992); (e) Suntzeff, Kinman \& Kraft (1994); (f) Walker \&
Terndrup (1991)}
\end{table}

\noindent
\begin{table}[t]
\caption{H and Ca spectral types (cols. 2 and 3), \ds ~values, and
corresponding metallicities for the program stars.
\label{tab-rrd2}}
\vspace*{5mm}
\begin{tabular}{lccccccccccc}
\tableline
\tableline
Star  &SpT &SpT &\ds   &\ds    &\ds   &[Fe/H] &[Fe/H] &[Fe/H]  &[Fe/H] &[Fe/H] 
&[Fe/H]\\
      &(H) &(K) & Kem  & T Sex &adopt. & Kem   & T Sex &adopt.    & Kem   & T 
Sex &adopt.\\
\multicolumn{6}{c}{} &\multicolumn{3}{c}{(Clementini et al. 1995)} 
&\multicolumn{3}{c}{(Gratton 1999)} \\
\tableline
CA 02 & F1.6 & A3.6 &  8.0 &  9.1 &  8.6 & $-$1.63 & --1.85 &--1.74 &$-$1.44 
&$-$1.63 &$-$1.54 \\
CB 45 & F1.9 & A4.2 &  7.5 &  8.4 &  8.0 & $-$1.54 &--1.71  &--1.62 &$-$1.35 
&$-$1.51 &$-$1.43 \\
CA 48 & A7.2 & A5.7 &  1.5 &  5.2 &  5.2 &         & --1.09 &--1.09 &        
&$-$0.95 &$-$0.95 \\
      & F2.2 & A7.3 &  4.9 &  5.7 &  5.3 & $-$1.03 & --1.19 &--1.11 &$-$0.89 
&$-$1.03 &$-$0.96 \\
CB 49 & F0.9 & A4.3 &  6.6 &  8.1 &  7.4 & $-$1.36 & --1.65 &--1.50 &$-$1.19 
&$-$1.46 &$-$1.33 \\
CB 61 & F2.7 & A6.8 &  5.9 &  6.4 &  6.2 & $-$1.14 & --1.32 &--1.23 &$-$1.07 
&$-$1.16 &$-$1.11 \\
CA 67 & F0.5 & A2.6 &  7.9 &  9.7 &  8.8 & $-$1.61 & --1.96 &--1.78 &$-$1.42 
&$-$1.74 &$-$1.58 \\
\tableline
\end{tabular}
\end{table}

\begin{table}[t]
\caption{Main parameters of the adopted pulsation models: ONLY
AVAILABLE IN ELECTRONIC FORM} 
\label{tab-mod} 
\end{table}

\noindent
\begin{table}[t]
{\footnotesize
\caption{\ds ~values, metal abundances on the two metallicity scales discussed
in the text, and individual masses for field RRd's. Metallicities for Sculptor 
and
Draco are for the whole systems, not for individual RRd's, and are given only 
once.
\label{tab-rrd3}}
\vspace*{5mm}
\begin{tabular}{lcccccc}
\tableline
\tableline
Star/cluster  &\ds &[Fe/H] &[Fe/H] &Mass (C95)  &Mass (G99)  &Reference\\
              &    & C95   &  G99  & M$_\odot$  & M$_\odot$  &(for ~$\Delta 
$S)\\
\tableline
LMC-CA 02   &8.6 &$-$1.74$\pm0.20$  &$-$1.54$\pm0.20$ & 0.588$\pm$0.077 & 
0.691$\pm$0.128 &(a)\\
LMC-CB 45   &8.0 &$-$1.62$\pm0.20$  &$-$1.43$\pm0.20$ & 0.742$\pm$0.135 & 
0.890$\pm$0.195 &(a)\\ 
LMC-CA 48   &5.2 &$-$1.09$\pm0.20$  &$-$0.95$\pm0.20$ & 1.034$\pm$0.240 & 
1.034$\pm$0.240 &(a)\\ 
LMC-CB 49   &7.4 &$-$1.50$\pm0.20$  &$-$1.33$\pm0.20$ & 0.860$\pm$0.240 & 
1.032$\pm$0.240 &(a)\\ 
LMC-CB 61   &6.2 &$-$1.23$\pm0.20$  &$-$1.11$\pm0.20$ & 1.075$\pm$0.250 & 
1.075$\pm$0.250 &(a)\\ 
LMC-CA 67   &8.8 &$-$1.78$\pm0.20$  &$-$1.58$\pm0.20$ & 0.765$\pm$0.102 & 
0.928$\pm$0.196 &(a)\\ 
MW-CU Com   &--  &$-$2.38$\pm0.20$  &$-$2.38$\pm0.20$ & 0.835$\pm$0.120 & 
0.835$\pm$0.120 &(b)\\ 
MW-VIII-10  &9.2&$-$1.86$\pm0.20$   &$-$1.66$\pm0.20$ & 0.708$\pm$0.063 & 
0.812$\pm$0.143 &(c)\\ 
MW-VIII-58  &8.6&$-$1.75$\pm0.20$   &$-$1.55$\pm0.20$ & 0.605$\pm$0.075 & 
0.722$\pm$0.138 &(c)\\ 
MW-AQ Leo   &8.9&$-$1.81$\pm0.20$   &$-$1.60$\pm0.20$ & 0.831$\pm$0.098 & 
0.987$\pm$0.199 &(d,c)\\ 
Draco-V11   &-- &$-$2.0$\pm0.15$    &$-$1.83$\pm0.15$ & 0.758$\pm$0.042 & 
0.786$\pm$0.054  &\\ 
Draco-V72   &-- &                   &                 & 0.763$\pm$0.008 & 
0.801$\pm$0.085  &\\ 
Draco-V83   &-- &                   &                 & 0.759$\pm$0.008 & 
0.798$\pm$0.086  &\\ 
Draco-V112  &-- &                   &                 & 0.810$\pm$0.008 & 
0.849$\pm$0.088  &\\ 
Draco-V138  &-- &                   &                 & 0.760$\pm$0.008 & 
0.798$\pm$0.084  &\\ 
Draco-V143  &-- &                   &                 & 0.783$\pm$0.007 & 
0.824$\pm$0.091  &\\ 
Draco-V156  &-- &                   &                 & 0.744$\pm$0.008 & 
0.781$\pm$0.080  &\\ 
Draco-V165  &-- &                   &                 & 0.634$\pm$0.009 & 
0.667$\pm$0.071  &\\ 
Draco-V169  &-- &                   &                 & 0.757$\pm$0.008 & 
0.795$\pm$0.085  &\\ 
Draco-V190  &-- &                   &                 & 0.751$\pm$0.008 & 
0.789$\pm$0.085  &\\ 
Sculptor-V1168 &-- &$-$1.8$\pm0.10$ &$-$1.58$\pm0.15$ & 0.782$\pm$0.092 & 
0.934$\pm$0.185  &\\ 
\tableline
\end{tabular}
\tablerefs{(a) This paper; (b) Clementini et al (2000a); (c) Clement et al.
(1991); (d) Mendes De Oliveira \& Smith (1990)}
}
\end{table}

\newpage

\noindent
\begin{table}[t]
{\footnotesize
\caption{\ds ~values, masses and metal abundances for the cluster RRd's.
Metallicities are for the whole systems, not for individual RRd's, and are
given only once per cluster.
\tablenotemark{1}
\label{tab-rrd4}}
\vspace*{5mm}
\begin{tabular}{lcccccc}
\tableline
\tableline
\\
Star/cluster  &\ds &[Fe/H] &[Fe/H] &Mass (C95)  &Mass (G99) &Reference\\
              &    &ZW84   & CG97  & M$_\odot$  & M$_\odot$ &(for ~$\Delta $S)\\
\tableline
NGC 6426 - V3 &--   &$-$2.20$\pm0.20$ &$-$2.11$\pm0.20$ & 0.814$\pm$0.039 & 
0.803$\pm$0.012   & \\
M 15 - V17    &11.3 &$-$2.15$\pm$0.20 &$-$2.12$\pm0.06$ & 0.849$\pm$0.072 & 
0.839$\pm$0.056   & (a)\\
M 15 - V26    &11.3 &                &                & 0.789$\pm$0.067 & 
0.780$\pm$0.052   & (a)\\
M 15 - V30    &11.3 &                &                & 0.776$\pm$0.063 & 
0.767$\pm$0.048   & (a)\\
M 15 - V31    &11.3 &                &                & 0.797$\pm$0.067 & 
0.787$\pm$0.052   & (a)\\
M 15 - V39    &11.3 &                &                & 0.790$\pm$0.072 & 
0.780$\pm$0.056   & (a)\\
M 15 - V51    &11.3 &                &                & 0.700$\pm$0.048 & 
0.693$\pm$0.036   & (a)\\
M 15 - V53    &11.3 &                &                & 0.803$\pm$0.066 & 
0.793$\pm$0.051  & (a)\\
M 15 - V58    &11.3 &                &                & 0.795$\pm$0.067 & 
0.786$\pm$0.052   & (a)\\
M 15 - V61    &11.3 &                &                & 0.758$\pm$0.061 & 
0.750$\pm$0.047   & (a)\\
M 15 - V67    &11.3 &                &                & 0.787$\pm$0.066 & 
0.777$\pm$0.051   & (a)\\
M 15 - V96    &11.3 &                &                & 0.790$\pm$0.070 & 
0.781$\pm$0.054   & (a)\\
NGC 2419 - V39 &--  &$-$2.10$\pm0.20$ &$-$1.97$\pm0.20$ &0.793$\pm$0.009 & 
0.801$\pm$0.030   & \\
M 68 - V7     &10.8 &$-$2.09$\pm0.20$ &$-$1.99$\pm0.06$ & 0.777$\pm$0.042 & 
0.769$\pm$0.011   & (a)\\
M 68 - V8     &10.8 &                &                & 0.761$\pm$0.039 & 
0.754$\pm$0.012   & (a)\\
M 68 - V3     &10.8 &                &                & 0.759$\pm$0.038 & 
0.752$\pm$0.012   & (a)\\
M 68 - V45    &10.8 &                &                & 0.746$\pm$0.036 & 
0.740$\pm$0.012   & (a)\\
M 68 - V19    &10.8 &                &                & 0.771$\pm$0.040 & 
0.764$\pm$0.011   & (a)\\
M 68 - V29    &10.8 &                &                & 0.768$\pm$0.039 & 
0.761$\pm$0.012   & (a)\\
M 68 - V4     &10.8 &                &                & 0.757$\pm$0.037 & 
0.750$\pm$0.012   & (a)\\
M 68 - V31    &10.8 &                &                & 0.765$\pm$0.037 & 
0.758$\pm$0.012   & (a)\\
M 68 - V34    &10.8 &                &                & 0.740$\pm$0.033 & 
0.735$\pm$0.012   & (a)\\
M 68 - V21    &10.8 &                &                & 0.772$\pm$0.037 & 
0.765$\pm$0.012   & (a)\\
M 68 - V26    &10.8 &                &                & 0.725$\pm$0.029 & 
0.721$\pm$0.012   & (a)\\
M 68 - V36    &10.8 &                &                & 0.717$\pm$0.027 & 
0.713$\pm$0.013   & (a)\\
\tableline
\end{tabular}
}
\end{table}

\addtocounter{table}{-1}

\begin{table}[t]
{\footnotesize
\caption{(continued)}
\vspace*{5mm}
\begin{tabular}{lcccccc}
\tableline
\tableline
\\
Star/cluster  &\ds &[Fe/H] &[Fe/H] &Mass (C95)  &Mass (G99)  &Reference\\
              &    &ZW84   & CG97  & M$_\odot$  & M$_\odot$  &(for ~$\Delta 
$S)\\
\tableline
M 3 - V68      &8.4  &$-$1.66$\pm0.20$ &$-$1.34$\pm0.06$ & 0.677$\pm$0.108 & 
0.930$\pm$0.204   & (a)\\
M 3 - V79      &8.4  &               &               & 0.772$\pm$0.140 & 
1.106$\pm$0.270   & (a)\\
M 3 - V87      &8.4  &               &               & 0.795$\pm$0.148 & 
1.147$\pm$0.286   & (a)\\
M 3 - V99      &8.4  &               &               & 0.846$\pm$0.164 & 
1.239$\pm$0.318   & (a)\\
M 3 - V166     &8.4  &               &               & 0.803$\pm$0.150 & 
1.160$\pm$0.289   & (a)\\
IC 4499 - V78  &6.9  &$-$1.50$\pm$0.20 &$-$1.26$\pm0.20$ & 0.841$\pm$0.181 & 
1.116$\pm$0.267   & (b)\\
IC 4499 - V18  &6.9  &               &               & 0.830$\pm$0.177 & 
1.100$\pm$0.260   & (b)\\
IC 4499 - V65  &6.9  &               &               & 0.832$\pm$0.177 & 
1.101$\pm$0.261   & (b)\\
IC 4499 - V10  &6.9  &               &               & 0.827$\pm$0.175 & 
1.093$\pm$0.257   & (b)\\
IC 4499 - V21  &6.9  &               &               & 0.833$\pm$0.177 & 
1.103$\pm$0.261   & (b)\\
IC 4499 - V51  &6.9  &               &               & 0.844$\pm$0.180 & 
1.118$\pm$0.266   & (b)\\
IC 4499 - V109 &6.9  &               &               & 0.850$\pm$0.183 & 
1.128$\pm$0.269   & (b)\\
IC 4499 - V59  &6.9  &               &               & 0.849$\pm$0.181 & 
1.125$\pm$0.267   & (b)\\
IC 4499 - V87  &6.9  &               &               & 0.856$\pm$0.184 & 
1.135$\pm$0.271   & (b)\\
IC 4499 - V63  &6.9  &               &               & 0.838$\pm$0.177 & 
1.107$\pm$0.260   & (b)\\
IC 4499 - V73  &6.9  &               &               & 0.854$\pm$0.182 & 
1.130$\pm$0.268   & (b)\\
IC 4499 - V42  &6.9  &               &               & 0.901$\pm$0.199 & 
1.204$\pm$0.294   & (b)\\
IC 4499 - V31  &6.9  &               &               & 0.855$\pm$0.181 & 
1.130$\pm$0.267   & (b)\\
IC 4499 - V90  &6.9  &               &               & 0.865$\pm$0.184 & 
1.144$\pm$0.271   & (b)\\
IC 4499 - V8   &6.9  &               &               & 0.887$\pm$0.190 & 
1.176$\pm$0.281   & (b)\\
IC 4499 - V71  &6.9  &               &               & 0.892$\pm$0.191 & 
1.182$\pm$0.282   & (b)\\
\tableline
\end{tabular}
\tablenotetext{1}{14 RRd's have been detected in M15, but 3 were not used
because their periods were given as uncertain. 
IC4499 contains 17 RRd's, but periods are available only for 16 of them.}
\tablerefs{(a) Costar \& Smith (1988); (b) Smith \& Perkins (1992)}
}
\end{table}

\newpage

\noindent
\begin{table}[t]
\caption{Mean masses for cluster and field RRd's.
\label{tab-rrd5}}
\vspace*{5mm}
\begin{tabular}{ccc}
\tableline
\tableline
 & ${\rm <M>_{GC}}$  &${\rm <M>_{field}}$ \\
\tableline
\multicolumn{3}{c}{ZW84} \\
${\rm [Fe/H] \le -1.7}$ & 0.772 ($\sigma$=0.032) & 0.744 ($\sigma$=0.072) \\
${\rm [Fe/H] > -1.7}$   & 0.835 ($\sigma$=0.048) & 0.928 ($\sigma$=0.155) \\
\multicolumn{3}{c}{CG97}\\
${\rm [Fe/H] \le -1.5}$ & 0.765 ($\sigma$=0.031) & 0.814 ($\sigma$=0.082) \\
${\rm [Fe/H] > -1.5}$   & 1.127 ($\sigma$=0.059) & 1.008 ($\sigma$=0.081) \\
\tableline
\end{tabular}
\end{table}

\vfill

\newpage
$$ $$
\newpage

\newpage

\begin{references}

\reference{} Alcock, C. {\it et al.} (the MACHO collaboration) 1996, \aj, 111,
 1146
\reference{} Alcock, C. {\it et al.} (the MACHO collaboration) 1997, \apj, 
 482, 89 (A97)
\reference{} Bertelli, G., Bressan, A., Chiosi, C., Fagotto, F., \& Nasi, E.
 1994, \aaps, 106,275
\reference{} Bono, G., Caputo, F., Castellani, V., \& Marconi M. 1996, \apj, 
 471, L33 (BCCM96)
\reference{} Bono, G., \& Stellingwerf, R.F. 1994, \apjs, 93, 233
\reference{} Bono, G., Caputo, F., Castellani, V., \& Marconi M. 1997a, 
 \aaps, 121, 327  
\reference{} Bono, G., Caputo, F., Cassisi, S., Incerpi, R., Marconi, M. 1997b, 
 \apj, 483, 811
\reference{} Butler, D. 1975, \apj, 200, 68
\reference{} Butler, D., \& Deming, D., 1979, \aj, 84, 86
\reference{} Carretta, E., \& Gratton R.G. 1997, \aaps, 121, 95 (CG97)
\reference{} Carretta, E., Gratton, R.G., \& Clementini G. 2000a, \mnras,
 316, 721
\reference{} Catelan, M. 1998, \apj, 495, L81
\reference{} Chadid, M., \& Gillet, D. 1998, \aaps, 260, 123
\reference{} Clari\'a, J.J, \&  Rosenzweig, P. 1978, \aj, 83, 278, (CR)
\reference{} Clement, C.M., Kinman, T.D., \& Suntzeff N.B. 1991, \apj, 372, 273
\reference{} Clement, C.M., Ferance, S \& Simon, N. 1993, \apj, 412, 183
\reference{} Clementini, G., Carretta, E,  Gratton, R.G., Merighi, R., Mould, 
 J.R., \& McCarthy, J.K. 1995, \aj, 110, 2319 (C95)
\reference{} Clementini, G., {\it et al.} 2000, \aj, 120, 2054
\reference{} Clementini, G.,  Gratton, R.G., Carretta, E., Bragaglia, A., 
 Di Fabrizio, L., \& Maio, M. 2001, \aj, submitted (C2001)  
\reference{} Clube, S.V.M., Evans \& Jones 1969, MemRAS, 72, 101
\reference{} Costar, D., \& Smith, H.A. 1988, \aj, 96, 1925 
\reference{} Corwin, T.M., Carney, B.W., \& Allen, D.M. 1999, \aj, 117, 1332
\reference{} Cox, A.N. 1991, \apj, 381, L71
\reference{} Cox, A.N. 1995,  in Astrophysical applications of powerful new 
 databases, S.J.Adelman, W.L.Wiese eds, Astronomical Society of the Pacific, 
 p. 243
\reference{} Crawford, D.L., 1975, \aj, 80, 955
\reference{} Crawford, D.L., 1978, \aj, 83, 48
\reference{} Crawford, D.L., 1979, \aj, 84, 1858
\reference{} Di Fabrizio, L. Clementini, G.,  Gratton, R.G., Carretta, E., \&
 Bragaglia, A. 2001, A\&A, submitted
\reference{} Eggen, O.J. 1994, \aj, 107, 1834
\reference{} Fernley, J.A., Skillen, I., \& Burki, G. 1993, \aaps, 97, 815
\reference{} Firmanyuk, B.N.,  Derevyagin,  V.G., \&  Lysova,  L.E. 1985,
 ATsir, 1374, 7
\reference{} Garcia-Melendo, E., \& Clement, C.M. 1997, \aj, 114, 1190
\reference{} Gillet, D., \& Crowe, R.A. 1998, \aap 199, 242
\reference{} Girardi, L., Bressan, A., Chiosi, C., Bertelli, G., \& Nasi, E.
 1996, \aaps, 117, 113
\reference{} Gratton, R.G. 1998, \mnras, 296, 739
\reference{} Gratton, R.G. 1999, in Globular Clusters, C.Martinez Roger,
 I.P\'erez Fourn\'on and F.S\'anchez eds, Cambridge University Press, p. 155
\reference{} Gratton, R.G., {\it et al.} 2001, \aap, in press
\reference{} Gratton, R.G., Fusi  Pecci, F., Carretta, E., Clementini, G.,
 Corsi, C.E., \&  Lattanzi, M. 1997, \apj, 491, 749
\reference{} Gratton, R.G., Sneden, C., Carretta, A., \& Bragaglia, A. 2000,
 \aap, 354, 169
\reference{} Gratton, R.G., Tornamb\'e, A., \& Ortolani, S. 1986, \aap, 169, 111
\reference{} Hauck, B., \& Mermilliod, M. 1998, \aaps, 129, 431
\reference{} Iglesias, C.A., \& Rogers, F.J. 1996, \apj, 464, 943
\reference{} Kaluzny, J., Kubiak, M., Szymanski, M., Udalski, A.,
 Krzeminski, W., \& Mateo, M. 1995, \aaps, 112, 407
\reference{} Kemper, E. 1982, \aj, 87, 1395
\reference{} King, J., Stephens A., \& Boesgaard, A.M. 1998, \aj, 115, 666
\reference{} Kinman, T.D., \& Carretta, E., 1992, \pasp, 104, 111
\reference{} Kovacs, G., Buchler, R., Marom , A. 1991, \aap, 252, L27
\reference{} Kovacs, G., Buchler, R., Marom , A., Iglesias, C. A., Rogers,
  F. J., 1992, \aap, 259, L46
\reference{} Kraft, R.P., Sneden, C., Langer, G.E., \& Prosser, C.F. 1992, 
 \aj, 104, 645
\reference{} Lub, J., 1979, Ph.D. Thesis
\reference{} Mateo, M., 1998, \araa, 36, 435
\reference{} Mendes de Oliveira, C., \& Smith, H.A., 1990, \pasp, 102, 652
\reference{} Nemec, J.M. 1985a, \aj, 90, 204
\reference{} Nemec, J.M. 1985b, \aj, 90, 240
\reference{} Oosterhoff, P.Th., 1939, Observatory, 62, 104
\reference{} Petersen, J.O. 1973, \aap, 27, 89
\reference{} Petersen, J.O. 1991, \aap, 243, 426
\reference{} Preston, G.W. 1959, \apj, 130, 507
\reference{} Preston, G.W., \& Paczyinski, B. 1964, \apj, 140, 181
\reference{} Rogers, F.J., \& Iglesias, C.A. 1992, \apj, 401, 361
\reference{} Sandage, A. 1993, \aj, 106, 703
\reference{} Smith, H.A. 1986, \pasp, 98, 1317
\reference{} Smith, H.A., \& Perkins, G.J. 1982, \apj, 261, 576 
\reference{} Suntzeff, N.B., Kinman, T.D., \& Kraft, R.P., 
 1994, \apjs, 93, 271
\reference{} Sweigart, A. 1997, \apj, 474, L23
\reference{} van Albada, T.S., \& Baker, N. 1971, \apj, 169, 311
\reference{} Walker, A.R. 1992, \apj, 390, L81
\reference{} Walker, A.R. 1994, \aj, 108, 555
\reference{} Walker, A.R., \& Nemec, J.M. 1996, \aj, 112, 2026
\reference{} Walker, A.R., \& Terndrup, D.M., 1991, \apj, 378, 119
\reference{} Zinn, R., \& West, M.J. 1984, \apjs, 55, 45

\end{references}
\end{document}